\documentclass[aps,twoside,prb,twocolumn,showpacs,showkeys,citeautoscript]{revtex4-1}
\usepackage{graphicx}
\usepackage{amsmath,amssymb,amstext}
\usepackage{datetime}
\usepackage{color}

\newcommand{\dd}{d}
\newcommand{\bvec}[1]{\boldsymbol{#1}}
\renewcommand{\vec}{\bvec}

\begin{document}

\title{Unconventional scaling of resistivity in two-dimensional Fermi liquids}

\author{Jonathan M. \surname{Buhmann}}
\affiliation{Institute for Theoretical Physics, ETH Z\"{u}rich, 8093 Z\"{u}rich, Switzerland}

\begin{abstract}
     We study the temperature dependence of the electrical resistivity of interacting two-dimensional Fermi liquids. We perform a numerical simulation of the nonequilibrium state based on semiclassical Boltzmann transport theory. A two-dimensional system of quasiparticles on a square lattice serves as a model system. We use a single-orbital tight-binding model of the dispersion. For conceptual purposes, we choose a simple repulsive on-site interaction that leads to quasiparticle scattering and delta-potential scatterers as a model of nonmagnetic impurity scattering. Through our simulation, we demonstrate that deviations from the predictions of standard Fermi-liquid theory can arise due to the nontrivial scattering geometry of umklapp processes, in special cases even in the ultra-low-temperature limit. We show through qualitative arguments how these unconventional scaling properties of the electrical resistivity, which are often interpreted as indication of a non-Fermi-liquid state, can arise due to special geometric conditions of the Fermi surface. The appearance of robust deviations from the predictions of Fermi-liquid theory within our simple model presents a novel viewpoint in order to interpret unconventional transport properties in electron-electron scattering dominated metallic systems.
          \begin{center}
     This paper is published as ''Phys. Rev. B \textbf{88}, 245128 (2013)''\\
     \url{http://link.aps.org/doi/10.1103/PhysRevB.88.245128}
     \end{center}
\end{abstract}

%\keywords{
%     mykeywords
%}

\maketitle

\section{Introduction}
\label{sec: introduction}

Probing the electrical resistivity and its temperature dependence is an important experimental approach to characterize a material. In particular, systems in which deviations from the expected standard Fermi-liquid behavior can be observed, generally receive great attention from both theoretical\cite{anderson_strange_2006,casey_hidden_2011,merino_transport_2000,deng_how_2013,maebashi_eletrical_1997,maebashi_eletrical_1998,rosch_zero_2005,pal_effect_2012,maslov_resistivity_2011} and experimental groups\cite{hussey2009,mackenzie2001,doiron_Fermi-liquid_2003,kostic_non-Fermi-liquid_1998}.

The standard behavior of electrical resistance includes more than one mechanism. Matthiessen's rule states that these contributions additively form the total resistivity. In the spirit of this decoupling, the standard contributions to the total resistivity are a temperature independent term from impurity scattering (residual resistance), a quadratically scaling term from two-particle collisions (Fermi liquid), and a term from scattering of charge carriers off phonons which grows with $T^{5}$ (Bloch-Gr\"{u}neisen) below the Debye temperature, while above this scale, this contribution scales linearly.

In the low-temperature regime, phonon scattering can be neglected and the standard form of the electrical resistivity reduces to
\begin{align}
	\rho\approx \rho_{0} + \alpha_{2}T^{2},
\label{eq: general form of quadratic resistivity}
\end{align}
where $\rho_{0}$ is the residual resistivity. A qualitative explanation for $\rho\sim T^{2}$ in generic Fermi liquids can be found in many textbooks\cite{smith_transport_1989, abrikosovMetals}.

A quadratic temperature scaling is usually considered as an identifying signature of a Fermi-liquid ground state. However, the experimentally observed resistivity often does not scale quadratically and it is sometimes argued that such a non-quadratic temperature dependence is a signature of a non-Fermi-liquid ground state or of a so-called bad-metal phase\cite{kostic_non-Fermi-liquid_1998,doiron_Fermi-liquid_2003,stewart_non-Fermi-liquid_2001}. Some of the best known examples of deviations from the standard scaling are the optimally doped high-temperature superconducting cuprates\cite{hussey2009, hussey2011, mackenzie1996, wuyts_Resistivity_1996, ando_electronic_2004}, the heavy-fermion compounds near quantum critical points\cite{custers2003}, and the bilayer ruthenate Sr$_{3}$Ru$_{2}$O$_{7}$ at the metamagnetic transition\cite{mackenzie2001}. In all of these compounds, the resistivity shows a linear temperature dependence.

Due to the recurrent interest in unconventional scaling laws, many attempts to explain the appearance of $\rho\sim T^{x}$ with $x<2$ have been presented\cite{anderson_strange_2006,casey_hidden_2011,merino_transport_2000,deng_how_2013,maebashi_eletrical_1997,maebashi_eletrical_1998,rosch_zero_2005,pal_effect_2012,maslov_resistivity_2011}. Deviations from the standard Fermi-liquid predictions can be found in a two-band model with a light band and a heavy band and strong impurity scattering close to a quantum critical point\cite{maslov_resistivity_2011} or in a strongly correlated system at high temperature beyond the existence of a quasiparticle peak in the spectral function\cite{deng_how_2013}. In our study, we address unconventional scaling of the resistivity in a single band model of a metal with well-defined quasiparticles. Our numerical method enables us to go beyond a discussion of the resistivity in terms of a single relaxation time which, in the presence of anisotropy, is inadequate to describe the dissipation mechanism. Furthermore, our method allows us to compute the resistivity at finite temperatures because the full energy dependence of the Boltzmann equation is taken into account. In consequence, we study the temperature dependence of the resistivity without restriction to an asymptotic low-temperature behavior.

In a related study, we have used our method to show that the thermoelectric effect in simple models of correlated Fermi liquids can deviate from the predictions of standard Fermi-liquid theory\cite{buhmann_thermopower_2013}. This unconventional behavior is a consequence of strongly anisotropic steady-state distribution functions. This anisotropy can be caused by geometric constraints for two-particle scattering or by strongly angular-dependent quasiparticle velocities.

In this paper, we present a numerical simulation of the resistivity of a two-dimensional Fermi liquid on a lattice. Our calculation is based on a numerical solution of the Boltzmann equation, which takes the full angular and radial degrees of freedom of the scattering geometry, the Fermi surface, and the quasiparticle velocities into account. This method is designed to study the resistivity as a function of the temperature beyond the single relaxation time approximation. We show that strongly anisotropic quasiparticle scattering and special scattering geometries can lead to robust deviations from the predictions of Fermi-liquid theory and, in particular, that the temperature dependence of the resistivity can be strongly influenced by the band structure.

\section{A simple model of an interacting Fermi liquid on a lattice}
\label{sec: the model}

In order to highlight the universality of our results, we use a very basic model of a metal. We study the resistivity of a two-dimensional electron system on a square lattice, parametrized by a simple tight-binding model,
\begin{align}
	\varepsilon_{\bvec{k}} & = -2t(\cos k_{x}+ \cos k_{y}) + 4t' \cos k_{x}\cos k_{y}.
\label{eq: tight binding model}
\end{align}

In order to study correlation effects, we use a repulsive on-site coupling $U$. In a similar yet more applied study\cite{buhmann_numerical_2013}, we have used a functional renormalization group approach in order to compute renormalized effective scattering vertices. Here, we do not follow this approach because detailed quantitative description of a special material is not the goal of this paper. Disorder is simulated by isotropic delta-potential scatterers, which is a simple but adequate approach to model non-magnetic impurities. For simplicity, we neglect the spin degeneracy of the electrons.

The dispersion \eqref{eq: tight binding model} and the quasiparticle coupling are treated as phenomenological input in our model and, thus, already contain the renormalization effects due to the interactions of the bare particles. Matching our calculations to an experiment would correspond to adjusting the dispersion parameters in order to approximate the low-energy states which can be experimentally accessed by, e.g., angle-resolved-photoemission spectroscopy.

This study is based on semiclassical Boltzmann transport theory in the presence of an external electric field that drives the system out of equilibrium. The relaxation of the quasiparticle distribution is described by the collision integral, an integration over all possible scattering processes weighted by the scattering rate. The form of the collision integral is discussed in detail in Sec. \ref{sec:EstimatingTheTemperatureDependenceOfTheResistivity}.

We solve the Boltzmann equation numerically by calculating the collision integral with its full angular and radial dependence in a discretized momentum space. The discretization is chosen to match the curvature of the Fermi surface in order to study the low-energy sector of the Brillouin zone at high resolution. We emphasize that the full calculation of the collision integral is important in order to take the anisotropy of the scattering rates and the quasiparticle velocities into account. Our numerical approach is described in more detail in the Appendix of Ref. \onlinecite{buhmann_numerical_2013}.

\section{Origins of electrical resistance and Matthiessen's rule}
\label{sec: origins of electrical resistance}

The electric field perturbs the system into a nonequilibrium state by transferring momentum to the distribution of quasiparticles. For a steady-state configuration to be realized, momentum must thus be dissipated to the environment. 

\begin{figure}[t]
\begin{center}

\includegraphics[width=0.5\textwidth]{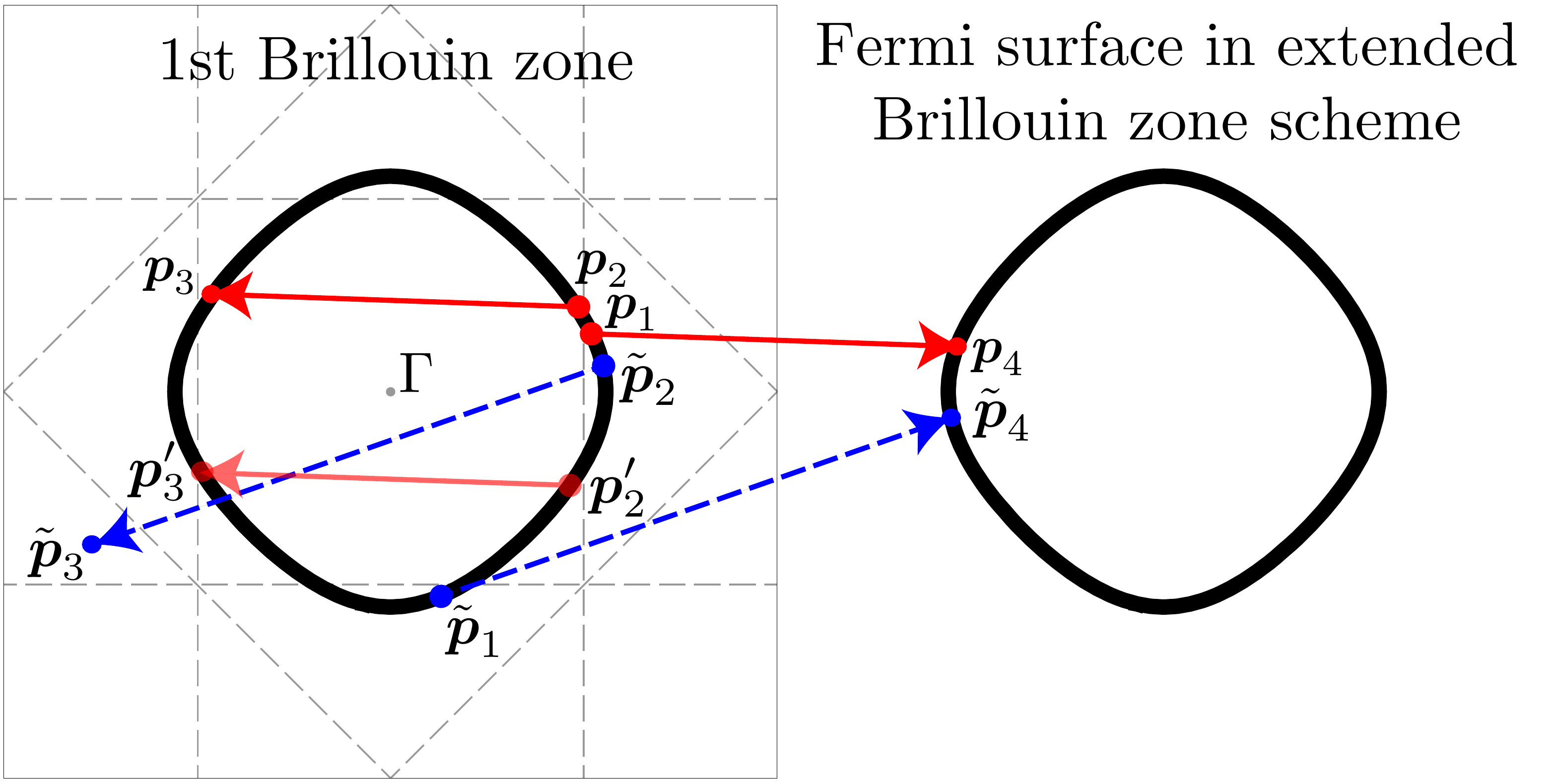}
\caption{Sketch of a typical two-particle umklapp scattering geometry: For the scattering process $\vec{p}_{1} \mapsto \vec{p}_{4}$, at zero temperature, there are only two possible scattering partners because all states must lie on the Fermi surface, $\vec{p}_{2}\mapsto \vec{p}_{3}$ or $\vec{p}'_{2}\mapsto \vec{p}'_{3}$. Umklapp scattering with $x$-component momentum transfer is not possible on the entire Fermi surface, as indicated by the scattering event sketched with dashed arrows. The minimal momentum transfer that connects the state $\tilde{\vec{p}}_{1}$ with the Fermi surface of a neighboring Brillouin zone is larger than the maximal momentum transfer between two states on the Fermi surface in the first Brillouin zone. Such a scattering event is forbidden by energy conservation, as $\varepsilon(\tilde{\vec{p}}_{3})\gg \varepsilon(\tilde{\vec{p}}_{1})+\varepsilon(\tilde{\vec{p}}_{2})-\varepsilon(\tilde{\vec{p}}_{4})$. The dashed gray lines are geometric borders for umklapp scattering, the umklapp lines.}
\label{fig: umklappEvent}

\end{center}
\end{figure}

One dissipative contribution arises from the interaction of the quasiparticles with lattice defects (impurity scattering), in which momentum is exchanged between the quasiparticles and the lattice. Two-particle scattering processes conserve the total momentum and, by itself, do not contribute to a finite resistivity. On a lattice, where the physical states live in the first Brillouin zone, quasiparticles can also scatter into a neighboring Brillouin zone. In terms of Bloch states, all final states have to be, however, defined in the first Brillouin zone. Final states in higher Brillouin zones are thus uniquely identified with states in the first Brillouin zone through a shift by a reciprocal lattice vector (cf. Fig. \ref{fig: umklappEvent}). Consequently, in the presence of a (crystal) lattice, only quasi-momentum is conserved and real momentum can be dissipated from the quasiparticle distribution to the lattice. This mechanism is referred to as umklapp scattering.

Umklapp scattering underlies strong geometrical constraints. In general, only states in the vicinity of the Fermi surface contribute to two-particle scattering in the low-energy limit. The vicinity of the Fermi surface, which we refer to as the set of \emph{thermally activated quasiparticle states}, can be characterized by the function $f^{0}(1-f^{0})$, where $f^{0}$ is the equilibrium distribution function. The width of this region and also its area is approximately linear in temperature. 

Due to the restriction of scattering states to a small region in momentum space, umklapp scattering can cause strong anisotropy. A simple example is sketched in Fig. \ref{fig: umklappEvent}, where umklapp scattering is possible in some regions of the Fermi surface (solid arrows) while for other parts of the Fermi surface, umklapp scattering with $x$-component momentum transfer is forbidden (dashed arrows).

If umklapp scattering is possible, this type of scattering will dominate the resistivity of clean systems. The normal scattering processes have been shown to contribute to the resistivity in the presence of disorder or in combination with umklapp scattering processes\cite{maebashi_eletrical_1998}. This coupling between scattering mechanisms represents a violation of Matthiessen's rule. In the following, we neglect the contribution of normal scattering processes for simplicity in our qualitative arguments. In the numerical calculations, however, all types of scattering are included and the effect of normal scattering processes is taken into account. In Ref. \onlinecite{buhmann_numerical_2013}, we have studied the violation of Matthiessen's rule with respect to impurity scattering and quasiparticle scattering.

\section{Estimating the temperature dependence of the resistivity}
\label{sec:EstimatingTheTemperatureDependenceOfTheResistivity}

We estimate the temperature dependence of the resistivity in terms of a very general argument for a generic Fermi surface shape. We neglect all the details of the band structure that are, however, crucial for a quantitative simulation. We only consider umklapp-scattering processes in order to estimate the resistivity scaling because this scattering mechanism gives the dominant contribution to the momentum relaxation.

Our scaling analysis is based on the temperature dependence of the nonequilibirum part $\delta f$ of the steady-state distribution function $f=f^{0}+\delta f$, the solution of the Boltzmann equation. The electrical current can be expressed in terms of $\delta f$ as
\begin{align}
    j_{x} = e \int \frac{\dd \vec{p}}{(2\pi\hbar)^{2}} \;v_{x}(\vec{p})\,\delta f(\vec{p}),
\label{eq:defOfElCurrent}
\end{align}
and, therefore, the temperature dependence of $\delta f$ is directly related to the temperature dependence of the electrical resistivity. We consider the left-hand side of the Boltzmann equation in the presence of an electric field
\begin{align}
	-e \vec{E}\cdot\vec{\nabla}_{p_{1}}f^{0} (\vec{p}_{1}).
\label{eq:driftTerm}
\end{align}
This drift term contains a factor of $\beta=1/T$ through $\vec{\nabla}f^{0}$ and hence $\beta$ also appears in $\delta f$. This factor $\beta$ is compensated in the definition of the electrical current [Eq. \eqref{eq:defOfElCurrent}], a momentum-space integral over $\delta f \propto f^{0}(1-f^{0})$. Hence, this integration of an area linear in temperature compensates for $\beta$ from the drift term \eqref{eq:driftTerm}. The remaining temperature dependence of $\delta f$ is determined by the right-hand side of the Boltzmann equation, the collision integral,
\begin{align}
	&-\int\frac{d \vec{p}_{2}}{(2\pi\hbar)^{2}}\,\frac{\dd \vec{p}_{3}}{(2\pi\hbar)^{2}}\,\frac{d \vec{p}_{4}}{(2\pi\hbar)^{2}}\;\Gamma(\vec{p}_{1},\vec{p}_{2},\vec{p}_{3},\vec{p}_{4})\nonumber\\ &\qquad\times\Bigl\{f(\vec{p}_{1})f(\vec{p}_{2})(1-f(\vec{p}_{3}))(1-f(\vec{p}_{4}))\nonumber\\
	&\qquad\qquad-(1-f(\vec{p}_{1}))(1-f(\vec{p}_{2}))f(\vec{p}_{3})f(\vec{p}_{4})\Bigr\}.\label{eq: e-e collision integral}
\end{align}

\begin{figure}[t]
\begin{center}

\qquad
\includegraphics[width=0.18\textwidth]{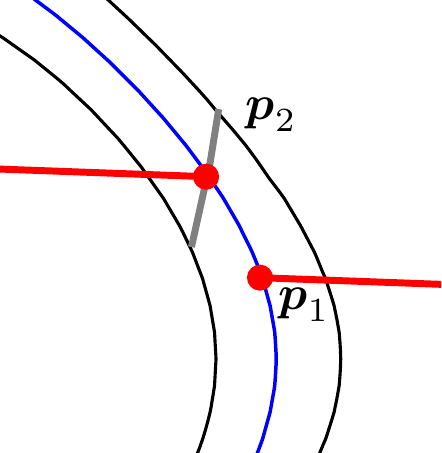}\hfill
\includegraphics[width=0.20\textwidth]{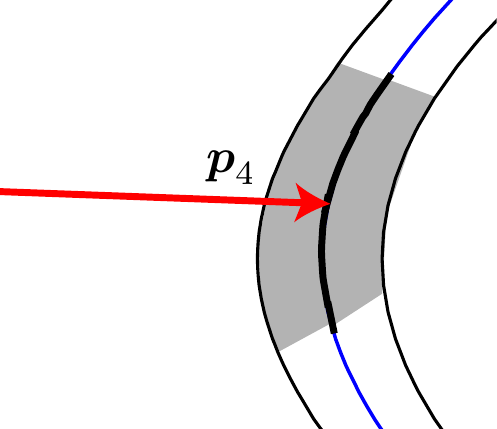}
\qquad
\phantom{.}
\caption{\emph{Left}: Freedom of the variable $\vec{p}_{2}$ in the generic scattering event of Fig. \ref{fig: umklappEvent} to yield a finite contribution in the collision integral. At zero temperature and fixed momentum transfer $\vec{p}_{1}-\vec{p}_{4}$, in this particular case exactly two states result in allowed final states; cf. Fig. \ref{fig: umklappEvent}. This constraint gives the integration variable $\vec{p}_{2}$ a zero-dimensional freedom. At finite temperature, for every $\vec{p}_{4}$ in the set of thermally activated quasiparticle states, the state $\vec{p}_{2}$ acquires a one-dimensional freedom. The reduction of a two-dimensional region of activates states around the zero-temperature state to one dimension is due to energy conservation. \emph{Right}: Freedom of the state $\vec{p}_{4}$. At zero temperature, all states on the Fermi surface are allowed (black line). At finite temperature, all thermally activated quasiparticle states around the Fermi surface yield possible final states (shaded area). This defines a two-dimensional area which scales linearly with temperature.}
\label{fig:TemperatureDependenceOfScatteringVolume}

\end{center}
\end{figure}

In two spatial dimensions, the collision integral consists of a six-dimensional integration. The scattering rates, derived via Fermi's golden rule from the microscopic interactions, ensure energy and momentum conservation,
\begin{align*}
	\Gamma(\vec{p}_{1},\vec{p}_{2},\vec{p}_{3},\vec{p}_{4})
	\propto &\ 
	\delta(\vec{p}_{1}+\vec{p}_{2}-\vec{p}_{3}-\vec{p}_{4})\nonumber\\&\ \times\delta(\varepsilon(\vec{p}_{1})+\varepsilon(\vec{p}_{2})-\varepsilon(\vec{p}_{3})-\varepsilon(\vec{p}_{4})).%\label{eq:ScatteringRatesFromFermisGoldenRule} 
\end{align*}
These delta functions reduce the dimensionality of the integration by two (momentum conservation) respectively by one dimension (energy conservation). Hence, the collision integral is effectively three-dimensional. The Fermi functions in Eq. \eqref{eq: e-e collision integral} further restrict the six-dimensional parameter space in which the kernel contributes to the integral. Thus, one can qualitatively estimate the temperature dependence of the resistivity by studying the temperature evolution of the volume in which the kernel takes a nonvanishing value. We refer to this volume as \emph{effective scattering volume}. It is important to only consider umklapp scattering events for the effective scattering volume, because this simple scaling analysis rests on the assumption that all scattering events contribute similarly to the relaxation of a charge current.

Figure \ref{fig: umklappEvent} displays a typical umklapp scattering process where $\vec{p}_{4}$ lies in a neighboring Brillouin zone. The six-dimensional integration can easily be reduced to four dimensions by evaluating the $\vec{p}_{3}$ integral in Eq. \eqref{eq: e-e collision integral} through the momentum-conserving delta function. This fixes the momentum $\vec{p}_{3}$ and, therefore, the temperature dependence of the effective scattering volume is characterized by the freedom of the variables $\vec{p}_{2}$ and $\vec{p}_{4}$. A detailed sketch of the parameter space for $\vec{p}_{2}$ and $\vec{p}_{4}$ is given in Fig. \ref{fig:TemperatureDependenceOfScatteringVolume}.

At zero temperature, the Pauli principle and the energy conservation restrict the four states $\vec{p}_{i}$ to the Fermi surface. Thus, focusing first on $\vec{p}_{4}$ (Fig. \ref{fig:TemperatureDependenceOfScatteringVolume}; right panel), this leaves a one-dimensional freedom at zero temperature. Every choice for $\vec{p}_{4}$ along the Fermi surface yields an allowed umklapp-scattering process. At finite temperature, the restriction of all states to the Fermi surface is softened. The states rather need to belong to the set of thermally activated quasiparticle states, which is bounded by a smooth momentum-space cutoff to the vicinity of the Fermi surface and which defines an area that is approximately linear in temperature. For the present qualitative analysis, one can interpret the cutoff for simplicity as a sharp cutoff, as indicated in Fig. \ref{fig:TemperatureDependenceOfScatteringVolume}. Thus, the finite-temperature freedom of the state $\vec{p}_{4}$ is increased to a two-dimensional area, which scales linearly in temperature due to the linear growth of the cutoff.

For any state $\vec{p}_{4}$, the energy and momentum transfer are fixed. At zero temperature, only two values of the integration variable $\vec{p}_{2}$ lead to a final state $\vec{p}_{3}$ on the Fermi surface (cf. Fig. \ref{fig: umklappEvent}). The product space of this zero-dimensional freedom with the one-dimensional freedom of the variable $\vec{p}_{4}$ defines one-dimensional parameter space with finite kernel at zero temperature. Thus, the effective scattering volume is a zero set and does not yield a finite contribution to the collision integral. Because the freedom of $\vec{p}_{4}$ grows linearly in temperature as a two-dimensional area, it could be expected that the freedom for $\vec{p}_{2}$ also grows in two dimensions. The momentum transfer is fixed, but at finite temperature, the energy difference between $\vec{p}_{2}$ and $\vec{p}_{3}$ varies for different choices of $\vec{p}_{2}$. Consequently, the freedom of $\vec{p}_{2}$ is reduced to the one-dimensional subspace defined by $\varepsilon(\vec{p}_{2})-\varepsilon(\vec{p}_{4})=\text{\it const}$. This subspace grows linearly with temperature, as sketched in Fig. \ref{fig:TemperatureDependenceOfScatteringVolume} (left panel). The freedom of the integration variables $\vec{p}_{2}$ and $\vec{p}_{4}$ results in a three-dimensional scattering volume, as required to obtain a finite contribution from the collision integral. The effective scattering volume scales with the temperature in two dimensions and thus shows an overall quadratic growth.

This simple argument explains the absence of a residual resistivity from two-particle scattering processes and, furthermore, illustrates that the natural temperature dependence of the resistivity of an interacting Fermi liquid is quadratic. It is important to realize that this argument is not derived from the quasiparticle lifetimes but from the scaling of the amount of umklapp processes that are in agreement with momentum and energy conservation and the Pauli principle. This argument requires only on very generic properties of a Fermi liquid and, thus, it is universal.

As a function of the band filling, the scattering geometry for umklapp scattering changes strongly. It is interesting to study special band-filling levels, for which a breakdown of our simple argument could be expected.

\section{Examples for band-structure-induced unconventional scaling}
\label{sec:ExamplesForUnconventionalScaling}

This study of the temperature scaling is based on numerical calculations of the resistivity as a function of the temperature for different band fillings within our two-dimensional square-lattice model. The temperature dependence of $\rho$ is analyzed using a power-law fit to the numerical data,
\begin{align}
	\rho(T)=\rho_{0}+\alpha T^{x},
\label{eq: power-law fit}
\end{align}
where the scaling exponent $x$ is a convenient quantitative measure of the resistivity scaling. 

In the remaining parts of this study, we discuss examples of unconventional temperature dependence of the resistivity by numerical simulations and qualitative explanations. We order these examples by their band filling. These special filling levels have already been described in our study of the thermoelectric effect, in which we have also investigated the properties of the electrical conductivity as a function of the chemical potential\cite{buhmann_thermopower_2013}.

\subsection{Below the onset of umklapp scattering}
\label{subsec: below U1}

\begin{figure}
\begin{center}

\includegraphics[width=0.45\textwidth]{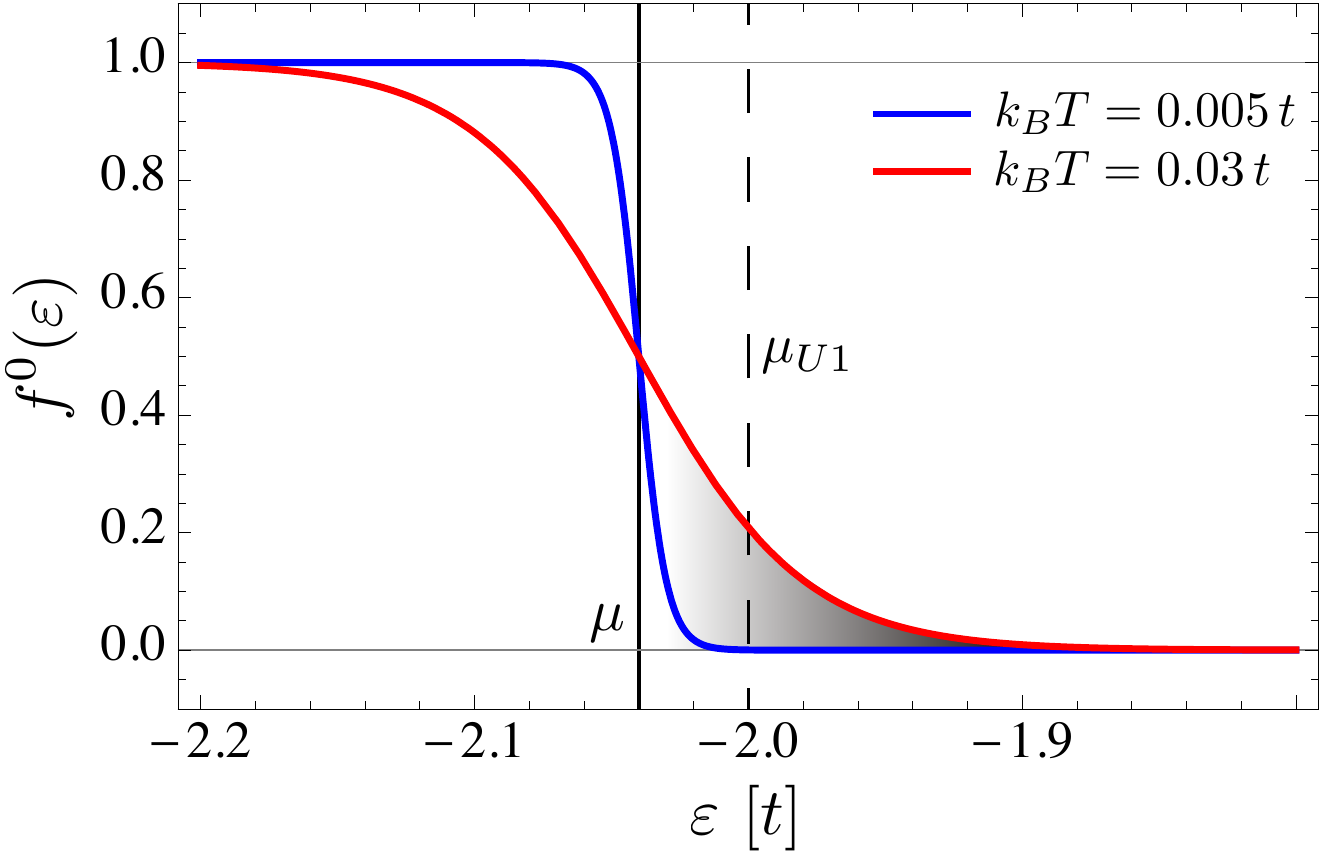}
\caption{Sketch of the thermally activated quasiparticle states that allow for umklapp scattering (shaded region) for band fillings slightly below the first umklapp edge. Two temperatures are compared. For $T<T_{\text{exp}}=0.02\,t/k_{B}$, umklapp scattering is exponentially suppressed due to the lack of thermally excited states above the umklapp edge. At temperatures above $T_{\text{exp}}$, thermally activated states allow for umklapp scattering. These two scenarios correspond to different temperature scaling of the resistivity (cf. Fig. \ref{fig: resistivityU1}).}
\label{fig:thermalActivation}

\end{center}
\end{figure}

\begin{figure*}
\begin{center}

\includegraphics[width=0.8\textwidth]{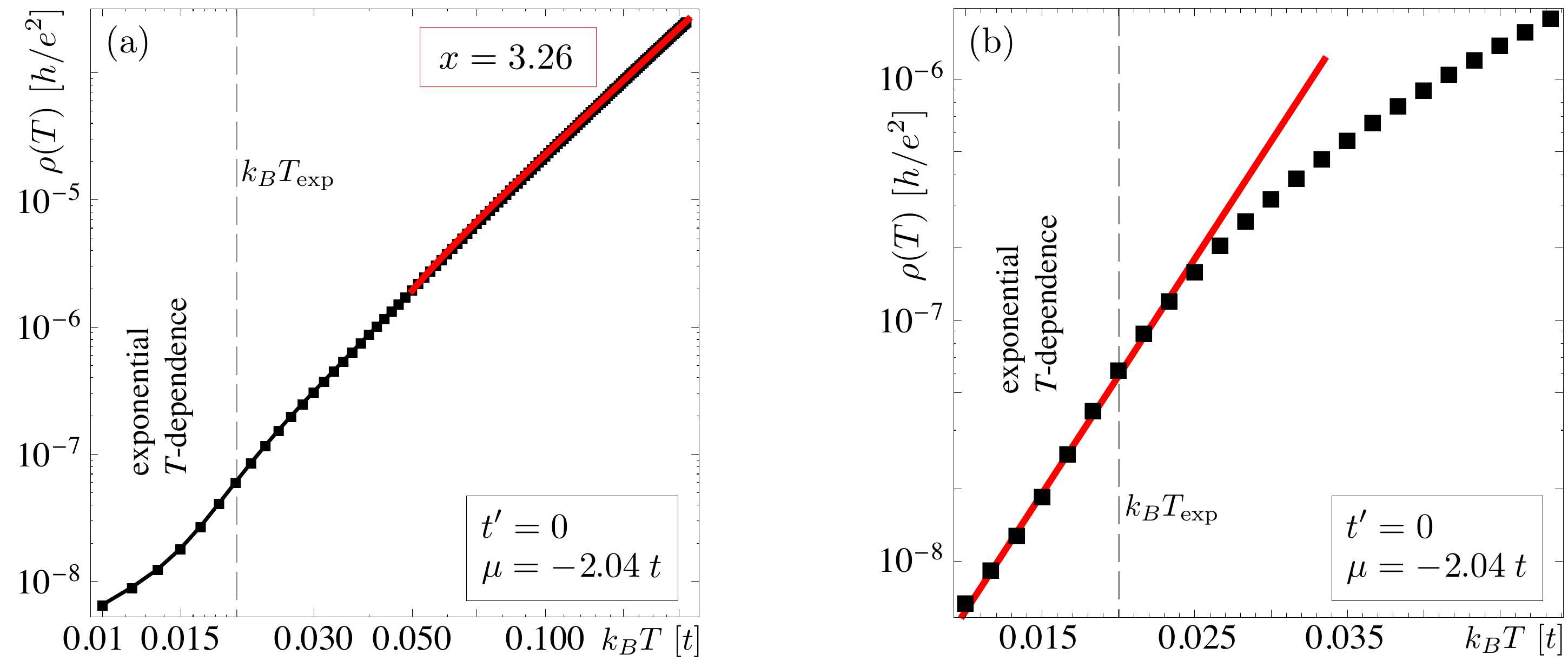}
\caption{\emph{Left}: A log-log plot of the resistivity is shown as a function of the temperature to illustrate non-power-law scaling at low temperature slightly below the first umklapp edge. \emph{Right}: A linear-log plot of the low-temperature regime of the resistivity is plotted. The linearity of $\rho$ on the logarithmic scale at low temperature illustrates the exponential growth of the resistivity. The vertical line represents the temperature scale $T_{\text{exp}}$ defined in the text as the crossover scale from exponential to non-exponential scaling.}
\label{fig: resistivityU1}

\end{center}
\end{figure*}

At very low band fillings, the Fermi surface is small and centered around the $\Gamma$-point of the Brillouin zone. There is a finite value of the chemical potential, $\mu=\mu_{U1}$, below which the Fermi surface is too small to contain states that match the energy and momentum requirements for umklapp scattering. Above $\mu_{U1}$, the Fermi surface crosses the vertical and horizontal umklapp lines (cf. Fig. \ref{fig: umklappEvent}), and umklapp scattering is possible. We refer to $\mu=\mu_{U1}$ as the first \emph{umklapp edge}, motivated by a strong decrease of the conductivity as a function of the band filling across the umklapp edge\cite{buhmann_thermopower_2013}. In our model [Eq. \eqref{eq: tight binding model}], the value of the chemical potential at the first umklapp edge is $\mu_{U1}=-2\,t$. 

The sharpness of the umklapp edge is true only at zero temperature. In the $T\rightarrow 0$ limit, only states on the Fermi surface contribute to a finite current and need to be relaxed. For finite temperature, the low-energy quasiparticle states in the vicinity of the Fermi surface also contribute to the nonequilibrium configuration. The energy scale that separates the high-energy states from the low-energy states is related to the temperature via the thermal broadening of the step in the Fermi function.

For band fillings sufficiently below the first umklapp edge, states that fulfill the criteria for umklapp scattering have exponentially small occupation numbers and, thus, their contribution to the resistivity is exponentially small (cf. Fig. \ref{fig:thermalActivation}). The resistivity scaling is based on the exponential growth of the occupation numbers and, thus, the natural temperature dependence from the scaling of the effective scattering volume as discussed in Sec. \ref{sec:EstimatingTheTemperatureDependenceOfTheResistivity} is suppressed. The temperature scale of this exponentially small growth is related to the energy difference between $\mu$ and $\mu_{U1}$ as long as $T\ll|\mu-\mu_{U1}|/k_{B}$. 

The temperature range of exponential growth of the resistivity ends with a crossover to a high-temperature regime with thermally activated quasiparticle states that fulfill the criteria for umklapp scattering. These states then dominate the temperature dependence of the resistivity and the scaling laws emerging from the umklapp-scattering processes are equivalent to those discussed in Sec. \ref{sec:EstimatingTheTemperatureDependenceOfTheResistivity}. An example of the resistivity slightly below the first umklapp edge is shown in Fig. \ref{fig: resistivityU1}, with $\mu=-2.04\,t<\mu_{U1}=-2\,t$. The crossover temperature from exponential to non exponential scaling in this simulation can be estimated as $k_{B} T_{\text{exp}}\approx (\mu_{U1}-\mu)/2 = 0.02\,t$, where the factor of 2 in the denominator represents the thermally broadened energy range of roughly $2k_{B}T$ above which high-energy states have exponentially small occupation numbers (cf. Fig. \ref{fig:thermalActivation}).

\subsection{Slightly above first umklapp edge}
\label{subsec: slightly above first umklapp edge}

For band fillings above the first umklapp edge, there are always states in the vicinity of the Fermi surface that match the umklapp-scattering conditions. The dominant contribution to the resistivity at low temperature originates from umklapp scattering of these states. In Sec. \ref{sec:EstimatingTheTemperatureDependenceOfTheResistivity}, it is explained that the natural temperature dependence of the contributions to the resistivity from regions of the Fermi surface that allow for umklapp scattering is quadratic. 

The magnitude of the resistivity can be estimated by the number of the states that satisfy the umklapp criteria. For band fillings only slightly above the umklapp edge, not all parts of the Fermi surface provide states that fulfill the geometric constraints for umklapp scattering (cf. Fig. \ref{fig: umklappEvent}). The size of the regions with umklapp scattering is temperature dependent. Due to thermal broadening, the minimal momentum transfer from within the range of thermally activated quasiparticle states to corresponding states in a neighboring Brillouin zone is decreased. Hence, combining the quadratic scaling with the temperature-driven enlargement of the Fermi-surface segments with umklapp scattering, we expect a non-power-law quadratic term with temperature-dependent coefficient slightly above the first umklapp.

In this case, a quadratic term survives in a zero-temperature Taylor series. This, however, does not imply that the resistivity follows a quadratic scaling for a finite-temperature interval. An example of $\rho(T)$ above the first umklapp edge is shown in Fig. \ref{fig: resistivityExamples} (a) with different power-law fits in the low- and the high-temperature regimes to illustrate that the resistivity does not follow a power-law $T$ dependence. The relaxation mechanism of systems slightly above the first umklapp edge is relatively similar to the high-temperature regime slightly below the umklapp edge.

\subsection{Close to the van Hove singularity}
\label{subsec:CloseToTheVanHoveSingularity}

It is well known, that scaling laws of transport quantities can be altered by the proximity to a singularity in the density of states\cite{si_linear_1991,hlubina_effect_1996,nakhmedov_density_2000,newns_quasiclassical_1994}. In our model of the dispersion [Eq. \eqref{eq: tight binding model}], saddle points at $(\pi/a,0)$ and $(0,\pi/a)$ cause a van Hove singularity. The effect on the scaling behavior of transport quantities is related to the unconventional growth of the number of thermally activated quasiparticle states, i.e., the discussion of Sec. \ref{sec:EstimatingTheTemperatureDependenceOfTheResistivity} and Fig. \ref{fig:TemperatureDependenceOfScatteringVolume} would not be valid for $\vec{p}_{2}$ or $\vec{p}_{3}$ close to a saddle point.

The two saddle points not only generate a large number of thermally activated states due to the flat dispersion, but they are also connected by the wave vector $(\pi/a,\pi/a)$, providing ideal conditions for umklapp scattering with  $(2\pi/a,2\pi/a)$ momentum transfer to the lattice. This strong scattering close to the saddle points induces strong anisotropy in the nonequilibrium distribution function. Another origin of anisotropy are the very slow quasiparticles close to the saddle points. We come back to another unusual feature related to the small quasiparticle velocities in the vicinity of a van Hove singularity in Sec. \ref{subsec: impurity minimum}.

Due to the strong anisotropy, it is generally not possible to give a universal description of the resistivity scaling of a system close to a van Hove singularity. In other words, even if we would be able to find a qualitative argument for a scaling law based on the number of thermally activated quasiparticle states close to the saddle points, one could not conclude from this analysis to the overall behavior. This insufficiency of an analysis based on a special selection of low-energy states can be interpreted as short circuiting of strong scattering contributions by quasiparticles with weak momentum relaxation\cite{hlubina1995}. The only qualitative statement that we can make is that our simple argument of Sec. \ref{sec:EstimatingTheTemperatureDependenceOfTheResistivity} breaks down. Thus, it is not straightforward to assume that the resistivity scales quadratically.

In Sec. \ref{sec:OverviewOfTheScalingExponent}, an overview of the scaling exponent as a function of the band filling is presented and in particular the region around the van Hove singularity is considered in detail. It is shown that the standard Fermi-liquid scaling behavior is modified in the vicinity of the van Hove singularity even at low temperature.

\subsection{Particle-hole symmetry at half-filling}
\label{subsec: perfect nesting}

\begin{figure}
\includegraphics[width=0.3\textwidth]{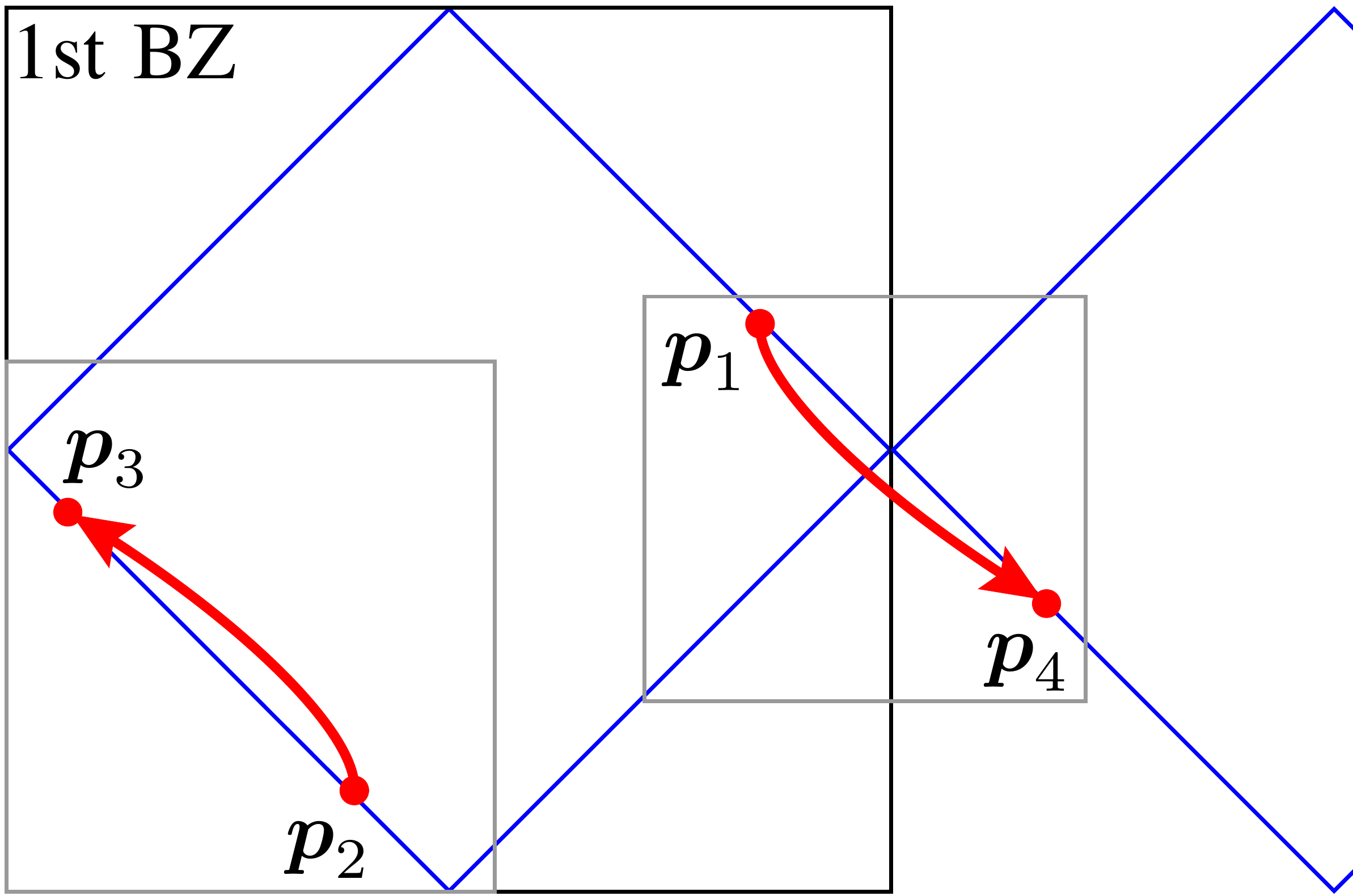}
\caption{Typical umklapp scattering process $(\vec{p}_{1},\vec{p}_{2})\mapsto(\vec{p}_{3},\vec{p}_{4})$ in a special nested Fermi-surface geometry. This scattering geometry leads to robust deviations from Fermi-liquid theory as is discussed in the text. The gray boxes around the integration parameters $\vec{p}_{2}$ and $\vec{p}_{4}$ mark the region in the Brillouin zone in which the scattering volume is analyzed (cf. Fig. \ref{fig:TemperatureDependenceOfScatteringVolumeNested}).}
\label{fig:TypicalUmklappEventNested}
\end{figure}

A very interesting scaling property can be qualitatively studied for the particle-hole-symmetric tight-binding model at half-filling ($t'=0;\;\mu=0$). In this scenario, the dispersion exhibits a diamond-shaped Fermi surface connecting the saddle points (cf. Fig. \ref{fig:TypicalUmklappEventNested}).

\begin{figure}[t]
\begin{center}

\quad\includegraphics[width=0.2\textwidth]{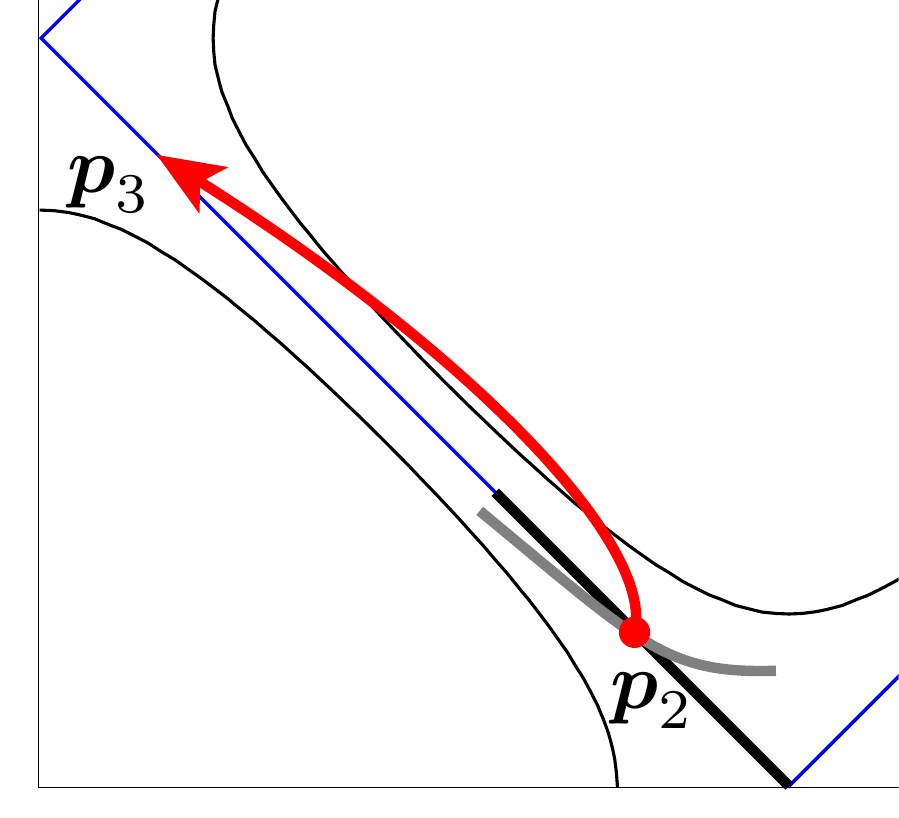}\hfill
\includegraphics[width=0.18\textwidth]{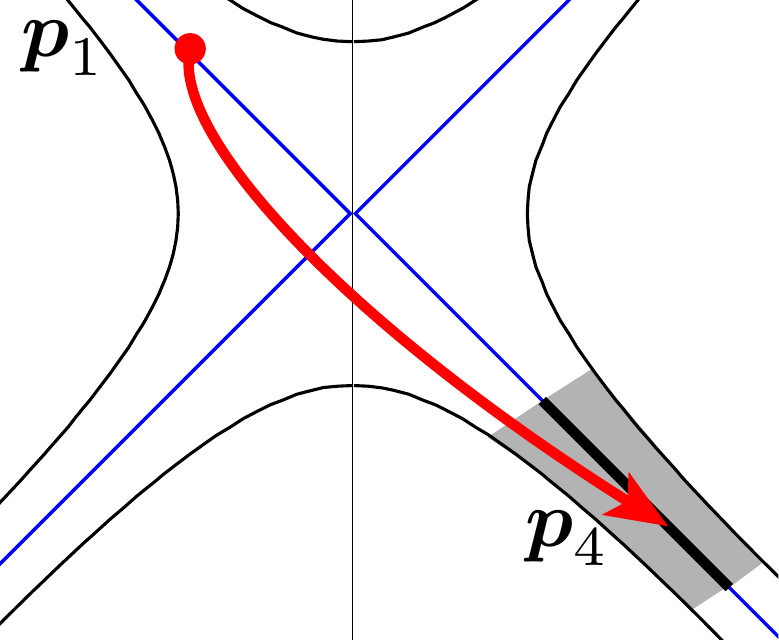}\quad\phantom{.}
\caption{\emph{Left}: Freedom of the state $\vec{p}_{2}$ for a typical scattering event in the special Fermi-surface geometry of Fig. \ref{fig:TypicalUmklappEventNested}. At zero temperature and fixed momentum transfer $\vec{p}_{1}-\vec{p}_{4}$, a large number of states $\vec{p}_{2}$ on the Fermi surface (black line) lead to scattering events which are in agreement with energy and momentum conservation. Thus, in contrast to the previous example, the integration variable $\vec{p}_{2}$ has a one-dimensional freedom at zero temperature. At finite temperature, for every state $\vec{p}_{4}$ in range of the thermally broadened Fermi surface (cf. right panel), the freedom of the state $\vec{p}_{2}$ remains a one dimensional due to the energy conservation. \emph{Right}: Freedom of the state $\vec{p}_{4}$. At zero temperature, states on the Fermi surface are allowed final states (black line). At finite temperature, states within a certain energy range around the Fermi surface are possible. This defines a two-dimensional volume, where the width scales linearly with temperature.}
\label{fig:TemperatureDependenceOfScatteringVolumeNested}

\end{center}
\end{figure}

In the zero-temperature limit, the final state $\vec{p}_{4}$ of the scattering process sketched in Fig. \ref{fig:TypicalUmklappEventNested} can be chosen freely along the Fermi surface, a one-dimensional region; cf. Fig. \ref{fig:TemperatureDependenceOfScatteringVolumeNested} (right panel). At finite temperature, the freedom is extended to a two-dimensional area in the Brillouin zone due to thermal broadening. The interesting difference to the generic case of Sec. \ref{sec:EstimatingTheTemperatureDependenceOfTheResistivity} is found considering the freedom of $\vec{p}_{2}$; cf. Fig. \ref{fig:TemperatureDependenceOfScatteringVolumeNested} (left panel). At zero temperature, the freedom of $\vec{p}_{2}$ has already a finite extension, i.e. one-dimensional, due to the nesting features of the Fermi surface. Increasing the temperature, the conservation of energy restricts the freedom at finite temperatures to one dimension, indicated by the gray line in Fig.  \ref{fig:TemperatureDependenceOfScatteringVolumeNested} (left panel). Summarizing the scaling analysis, at zero temperature the combined freedom of the integration variables $\vec{p}_{4}$ and $\vec{p}_{2}$ yields a two-dimensional scattering volume, a zero-set in the effective three-dimensional collision integral. At finite temperature, the enlarged range of the possible final states $\vec{p}_{4}$ leads to the required third dimension for a finite contribution from the collision integral. In this case, however, the growth is linear in temperature, because the freedom of $\vec{p}_{2}$ does not grow with temperature but remains one dimensional (cf. black and gray lines in Fig. \ref{fig:TemperatureDependenceOfScatteringVolumeNested}).

In addition to this particular scattering process that we have considered, there are a number of qualitatively different scattering events for this form of the Fermi surface, e.g., umklapp scattering with momentum transfer of $(2\pi/a,2\pi/a)$ to the lattice. For the majority of scattering processes, equivalent arguments lead to a linear growth of the scattering volume.

In summary, this simple scattering volume argument explains a linear temperature dependence of the resistivity for this kind of Fermi surface. Although this illustration relies on a very special geometry, it is remarkable because a linear temperature dependence of the resistivity represents a robust deviation from the predictions of Fermi-liquid theory.

This result is derived for a system that is associated with a van Hove singularity, but it is important to note that our argument does not rely on the flat dispersion around the saddle points or the peak in the density of states. The linear scaling is a consequence of this specially nested Fermi surface. Furthermore, in the zero-temperature limit, the above argument is expected to hold only for a system with exactly this Fermi surface. At higher temperature, it is reasonable to expect that an emerging linear term is present even for systems that approximately fulfill these special nesting conditions. In such a case, depending on the details of how good the nesting conditions are met, one can expect a competition between linear and quadratic terms.

\begin{figure*}
\begin{center}
\includegraphics[width=0.65\textwidth]{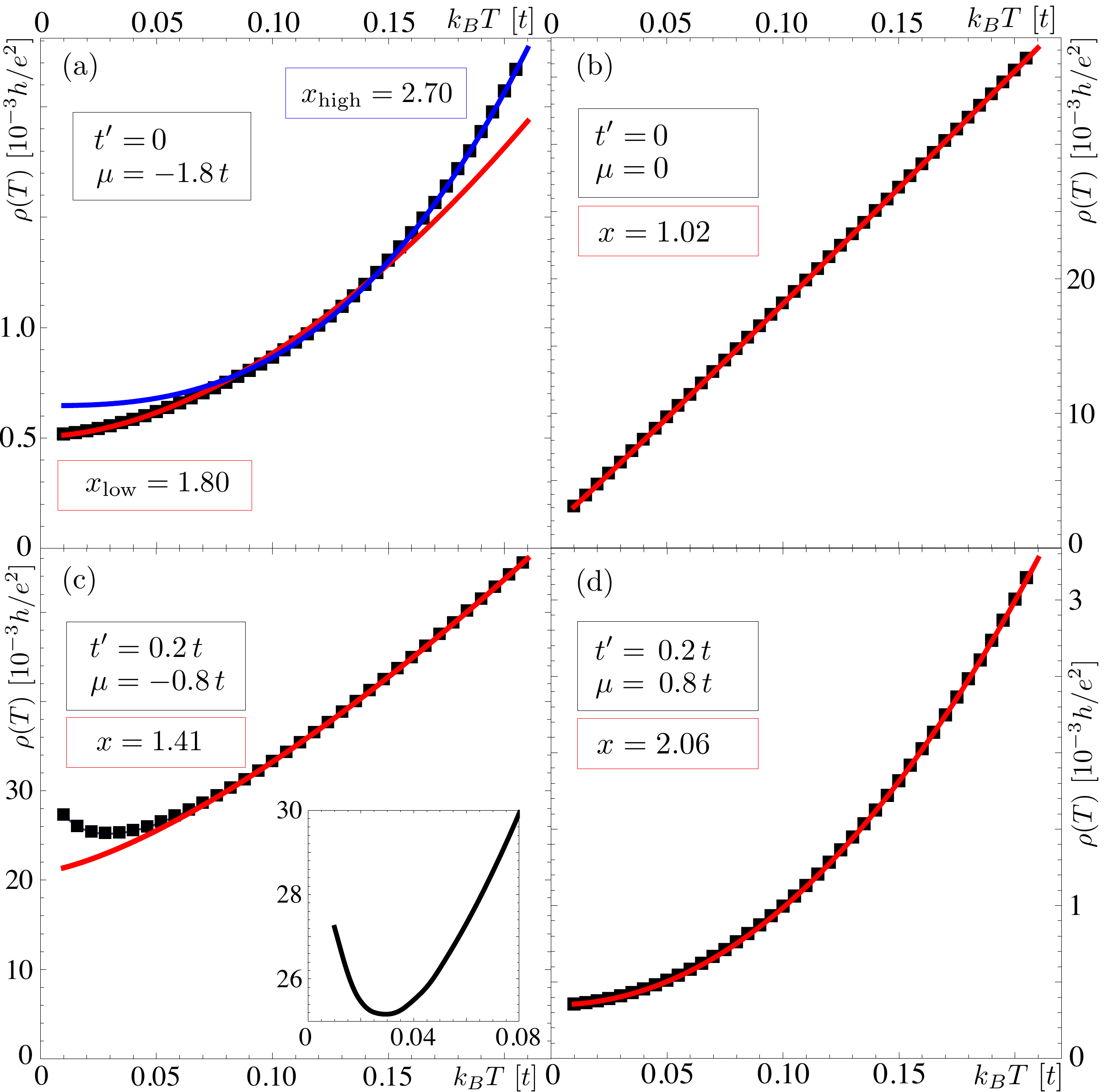}
\caption{Four examples of a non-universal temperature dependence of the resistivity in a temperature regime from $T=0.01\,t/k_{B}$ to $T= 0.2\,t/k_{B} $ (insets: band structure parameters and scaling exponents of a power-law fit): (a) $\rho(T)$ slightly above the first umklapp edge yielding non-power-law temperature dependence. (b) perfectly linear resistivity in the half-filled particle-hole-symmetric model. (c) resistivity minimum due to impurity scattering. (d) quadratic temperature dependence for a band filling far from van Hove singularity and umklapp edges.}
%\vspace{-10pt}
\label{fig: resistivityExamples}
\end{center}
\end{figure*}

In Fig. \ref{fig: resistivityExamples} (b), a simulation of the resistivity of the half-filled particle-hole-symmetric system is shown. This plot convincingly demonstrates the linearity of the resistivity over a wide range of temperature, down to the lowest temperature of our study.

It is important to point out the particular nesting conditions that are required to observe such a linear scaling of the effective scattering volume, which eventually leads to a linear resistivity. Following our above argument, we conclude that scattering processes with linear growth of the corresponding effective scattering volume are present if the Fermi surface exhibits segments that are parallel to one of the umklapp lines of Fig. \ref{fig: umklappEvent} and have a symmetric counterpart defined by a reflection about the umklapp line. One can even allow for a displacement by a parallel shift along this umklapp line. If this condition applies only for a small part of the Fermi surface, the majority of all scattering events can still lead to a normal, i.e. quadratic growth of the effective scattering volume. It depends on the exact form of the Fermi surface and requires a numerical simulation to extract the scaling behavior quantitatively; cf. the discussion about short-circuiting large scattering contributions in Sec. \ref{subsec:CloseToTheVanHoveSingularity}. It is, however, possible that such a nested Fermi-surface segment leads to a deviation from the ideal Fermi-liquid scaling. A special case is realized when the Fermi surface fully coincides with the umklapp lines, i.e. the above criterion is fulfilled for the entire Fermi surface. Under this condition, the umklapp scattering processes with linear growth of the effective scattering volume dominate the low-temperature transport and lead to a linear resistivity, as in the above example.

\subsection{Impurity induced resistivity minimum close to the van Hove singularity}
\label{subsec: impurity minimum}

So far, only the temperature dependence due to two-particle-scattering events has been discussed. In this section, we demonstrate how impurity scattering together with a strongly energy dependent density of states can also contribute to a nontrivial temperature dependence of the resistivity. 

In the presence of disorder, the divergence in the single-particle density of states is likely to be removed\cite{hlubina_effect_1996}. However, because the following qualitative argument does not depend on the divergence of the density of states, we neglect the softening of the van Hove singularity due to impurity scattering. 

We start with a qualitative discussion of the impurity-scattering-induced resistivity scaling in the vicinity of the van Hove singularity. For simplicity, we consider the single-relaxation-time approximation, which has proven to be a good description of isotropic impurity scattering,
\begin{align}
	\sigma^{\text{imp}}_{xx}\sim\tau\beta \int\dd \varepsilon N(\varepsilon) f_{0}(\varepsilon)(1-f_{0}(\varepsilon))v_{x}^{2}(\varepsilon).
\label{eq: conductivity due to impurities}
\end{align}

According to the above qualitative expression of the conductivity, one needs to compare the density of states with the inverse dispersion gradient, $N(\varepsilon)\sim (\partial\varepsilon/\partial k)^{-1}\sim v^{-1}(\varepsilon)$, in order to determine if the conductivity is enhanced or reduced by a singularity in the density of states. This argument leads to 
\begin{align}
	\sigma^{\text{imp}}_{xx} \sim \overline{v}_{x},
	\label{eq:SigmaSimVk}
\end{align}
where the average of the quasiparticle velocity $\overline{v}_{x}$ is to be taken over the thermally activated quasiparticle states. Equation \eqref{eq:SigmaSimVk} implies that a peak in the density of states leads to a peak in the resistivity which corresponds to a dip in the conductivity. 

In the vicinity of a peak in the density of states, the relative number of excited states that are very close to the saddle points depends on the temperature. While at low temperature, these states represent a substantial contribution in the average of Eq. \eqref{eq:SigmaSimVk}, at higher temperature, more normal states are thermally activated and the effect of the states close to the saddle points is effectively reduced. This implies that the contribution of the states with very flat dispersion and consequently low velocities is suppressed with increasing temperature. Thus, the conductivity increases with temperature, which represents a very unusual feature in the charge transport of metals.

Taking the quasiparticle interactions into account, for sufficiently large temperature, the effect of two-particle scattering dominates the effect of impurity scattering and leads to an increasing resistivity, resembling the conventional trend. A resistivity that decreases at low temperature and then finally grows, inevitably has a minimum at finite temperature.

The numerical simulation shown in Fig. \ref{fig: resistivityExamples} (c) demonstrates that the presence of strong impurity scattering can lead to a minimum in the resistivity as a function of the temperature for band fillings close to a peak in the density of states.

\subsection{Above the last umklapp edge}

For band fillings above the last umklapp edge, umklapp scattering eventually becomes fully suppressed. In this case, the same argument as given for systems below the first umklapp edge holds; cf.  Sec. \ref{subsec: below U1}. This can be easily understood in terms of a particle-hole transformation that relates almost filled bands to almost empty bands.

\subsection{Resistivity in three-dimensional metals}

In this discussion, we have presented intriguing examples of unconventional temperature dependencies in two-dimensional (layered) Fermi liquids. Naturally, the question rises as to whether similar examples can be found in fully three-dimensional systems. In principle, our dimensional arguments can be extended to three-dimensional models with a higher-dimensional integral in the Boltzmann equation and higher-dimensional effective scattering volume. It should be noted, however, that the generic case that results in conventional scaling of the resistivity is much more robust in three dimensions than in two-dimensional models.

\section{Power-law scaling exponent of the resistivity}
\label{sec:OverviewOfTheScalingExponent}

The scaling exponent $x$ [Eq. \eqref{eq: power-law fit}], can be used as a tool to characterize the growth of the resistivity. We analyze $x$ as a function of the chemical potential over the entire range in band filling.

\begin{figure}
\begin{center}

\includegraphics[width=0.4\textwidth]{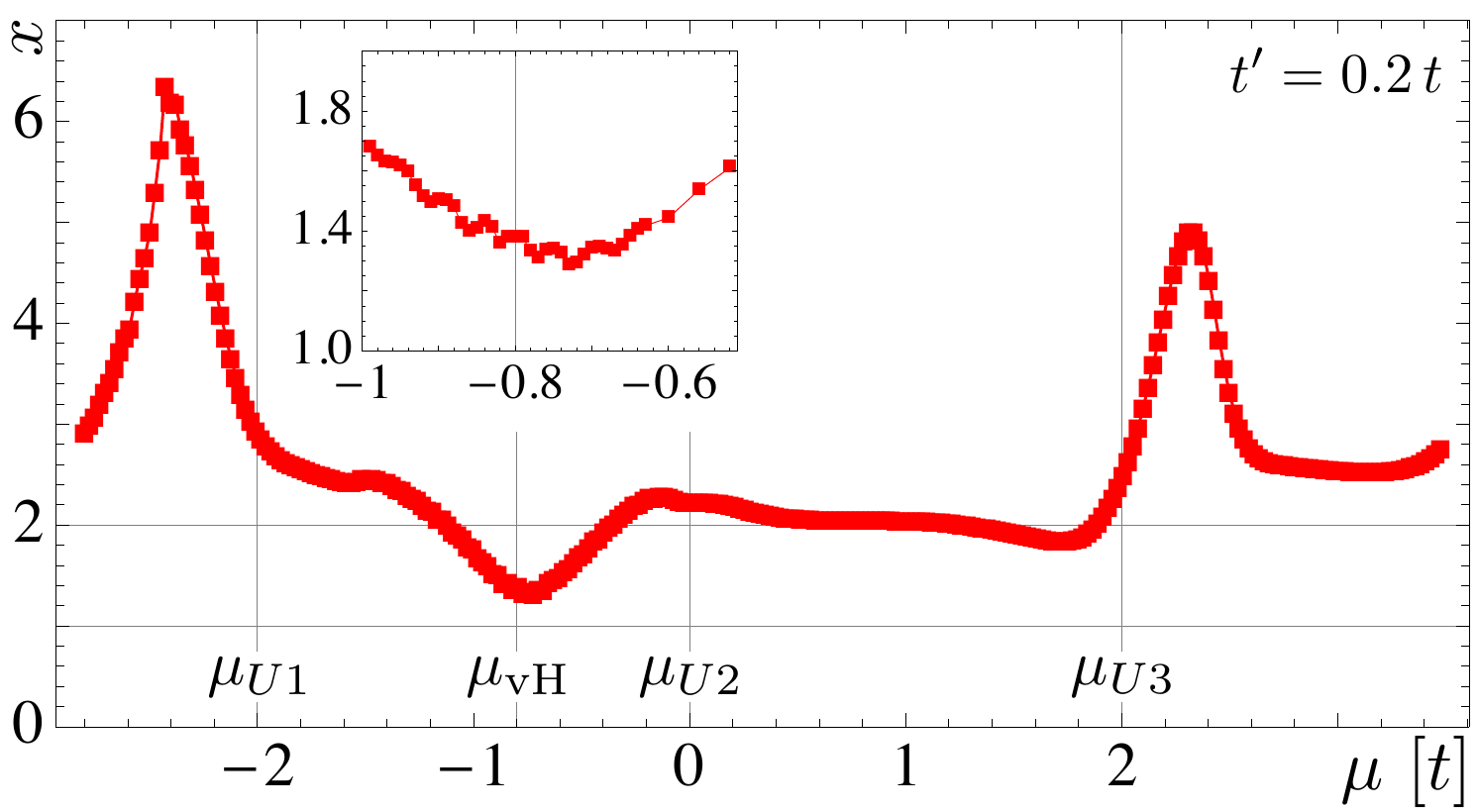}\\
\includegraphics[width=0.4\textwidth]{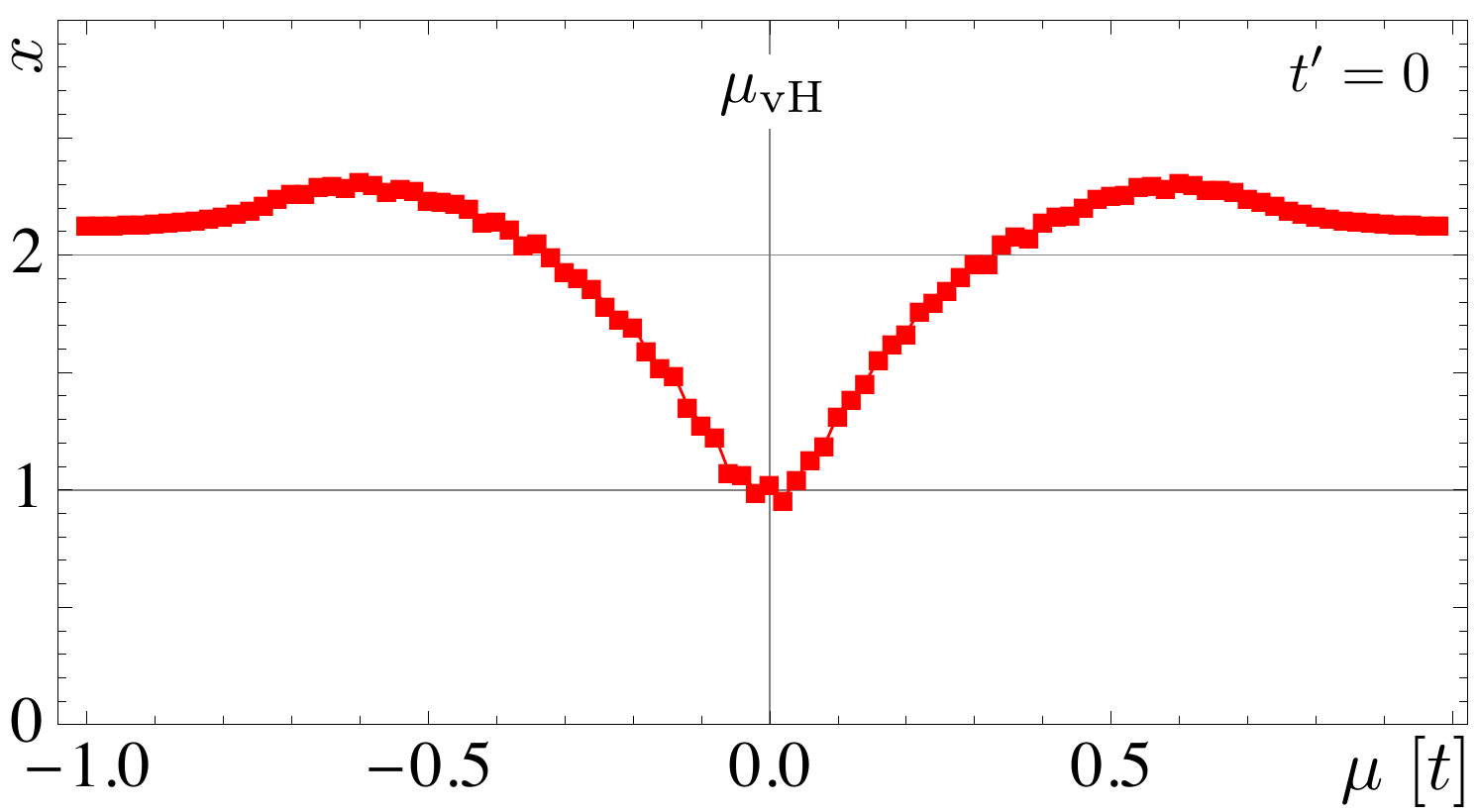}
\caption{\emph{Top}: Scaling exponent $x$ [Eq. \eqref{eq: power-law fit}] of the resistivity as a function of the band filling for finite $t'$. The peaks show the umklapp edges and are signatures of exponential temperature dependence. Note that for finite $t'$, even exactly at the van Hove singularity $\mu=\mu_{\text{vH}}$, the resistivity is not perfectly linear. \emph{Bottom}: Scaling exponent of the resistivity as a function of the chemical potential for a vanishing value of $t'=0$. In both cases, the fit is performed from $T=0.01\,t/k_{B}$ to $T=0.2\,t/k_{B}$.}
\label{fig: exponentBand}

\end{center}
\end{figure}

This analysis is based on the discussion of $x(\mu)$ for a generic choice of $t'=0.2$. The power-law fit to the numerical data is performed from $T=0.01\,t/k_{B}$ to $T=0.2\,t/k_{B}$. The scaling exponent $x$ is shown in Fig. \ref{fig: exponentBand} (upper panel). The first ($\mu_{U1}=-2\,t$) and last umklapp edge ($\mu_{U3}=2\,t$) can be clearly identified as peaks in the exponent. The peaks are a signature of the exponential temperature dependence outside of the umklapp regime, as discussed in Sec. \ref{subsec: below U1}. Around the van Hove filling, $x$ deviates downwards from $x>2$ to values substantially lower than two. In Secs. \ref{subsec:CloseToTheVanHoveSingularity} and \ref{subsec: perfect nesting}, we have explained how to understand this behavior. For the finite choice of $t'$, the scaling exponent does not reach $x=1$, (cf. inset of Fig. \ref{fig: exponentBand}; top panel). The reason for not reaching linear-$T$ is due to the finite value of $t'$ and the mismatch of the Fermi surface with the special nested Fermi surface of our example of Sec. \ref{subsec: perfect nesting}. The $x<2$ scaling in this example can be interpreted in terms of the nesting conditions which are only met approximately or the unconventional enhancement of the number of thermally activated quasiparticle states close to a van Hove singularity. The inset of this figure shows that the minimum of $x$ is slightly shifted away from the van Hove singularity, indicating that the unconventional trend is not only induced by the singularity in the density of states.

Figure \ref{fig: exponentBand} (bottom) displays the data from an equivalent calculation with vanishing $t'$ in the regime around the van Hove filling. In this case, as can be expected from Fig. \ref{fig: resistivityExamples} (b) and the discussion in Sec. \ref{subsec: perfect nesting}, the scaling exponent goes down to $x=1$ (linear resistance) at half-filling. 

It is interesting to note that the regime with $x<2$ in the vicinity of the van Hove filling is accompanied on both sides by two peak-like features. With an increase of the temperature, the system fulfills the nested scattering geometry at half-filling approximately, as explained in Sec. \ref{subsec: perfect nesting}. In this scattering geometry, not only the trend toward linearity can be understood, but also a large absolute value is expected from the amount of possible scattering processes [cf. the absolute value of $\rho$ in Figs. \ref{fig: resistivityExamples} (b) and (d)]. For systems that are close to this scattering configuration, the large scattering contribution is not present at lowest temperature due to the mismatch with the nested Fermi surface. With increased temperature, the nesting conditions are then roughly met due to thermal broadening. One can understand this as the opening of new scattering channel which becomes dominant with increasing temperature. The temperature-induced switching on of an additional large scattering mechanism leads to an upwards turn in the resistivity, which results in a locally increased curvature. Consequently, the exponent of the best power-law fit is expected to be enhanced.

In order to examine this feature in more detail, the so-called \emph{local scaling exponent} $x_{l}(T)$ can be studied. This $x_{l}(T)$ is the temperature-resolved analog of the scaling exponent in the above discussion. This temperature-dependent scaling exponent of the resistivity is defined for each temperature and it is formulated as
\begin{align}
	\rho(T) = \rho_{0} + \alpha T^{x_{l}},\qquad x_{l}=x_{l}(T),
\label{eq: definition of the local exponent}
\end{align}
with $x_{l}(T)$ obtained by taking the logarithmic derivative of $\rho(T)$,
\begin{align}
	x_{l}(T) = \frac{\dd}{\dd \ln T} \ln(\rho(T) - \rho_{0}),
\label{eq: logarithmic derivative of rho yields x}
\end{align}
with the residual resistivity $\rho_{0}$. 

\begin{figure}
%\begin{center}

\includegraphics[width=0.49\textwidth]{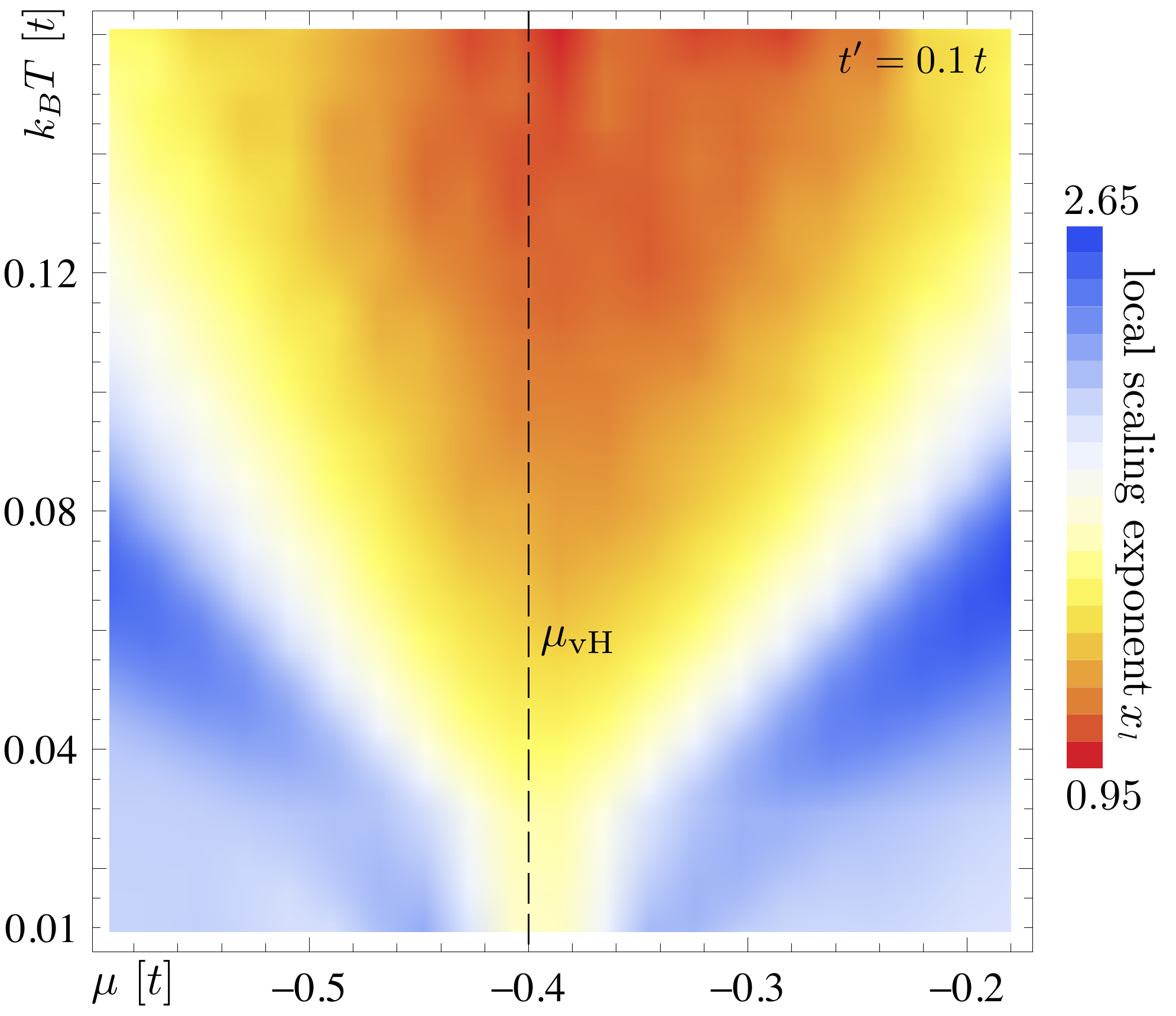}
\caption{Local scaling exponent $x_{l}$ [cf. Eq. \eqref{eq: logarithmic derivative of rho yields x}] of the resistivity as a function of the chemical potential and temperature for a finite value of $t'=0.1$. One clearly identifies the unconventional scaling ($x_{l}<2$) of the resistivity in the high-temperature regime above the van Hove filling at $\mu=-0.4\,t$.}
\label{fig: localExponent}

%\end{center}
\end{figure}

We calculate $x_{l}(T,\mu)$ over a finite range in band filling around the van Hove filling for a small but finite next-to-nearest-neighbor hopping parameter of $t'=0.1$. The results are shown as color scale in Fig. \ref{fig: localExponent}. One can clearly identify a regime where the scaling is almost linear (high-temperature around the van Hove level) and regimes where the scaling is mostly quadratic. With this temperature resolved analysis, one can identify that there is a crossover range from quadratic to linear at finite temperature. This crossover temperature scales apparently with the difference between the chemical potential and the energy scale of the van Hove level.

Consistent with the above interpretation of the scaling peaks on both sides of the regime with $x<2$ around the van Hove filling, $x_{l}$ acquires values exceeding two in between the linear and the quadratic scaling regimes. This large scaling exponent is related to the enhanced curvature due to the stronger slope of the linear contribution that only appears at higher temperature.

\section{Discussion and conclusions}
\label{sec: discussion and conclusions}

In this paper, we studied the temperature dependence of the electrical resistivity of a correlated two-dimensional Fermi liquid. For this purpose, numerical simulations of the resistivity within interacting tight-binding models on a square lattice have been performed and interpreted through qualitative arguments. In our model, quasiparticle interactions are given by a bare Hubbard-type on-site repulsion and impurities are modeled as delta-potential scatterers. This very simple model is chosen in order to emphasize the universality of the results and the potential relevance of our conclusions in multiple contexts.

We have demonstrated that the charge transport of simple metals in two dimensions has a much richer structure than is usually discussed in standard textbooks. While the resistivity due to electronic correlations is assumed to scale quadratically with temperature, it can be concluded from our simulation, that this is only true in the absence of special features in the band structures that influence the scattering geometry. Examples of such features are umklapp edges, van Hove singularities, or segments of the Fermi surface that fulfill special nesting requirements. In the ideal situation, the dominant umklapp-scattering processes can be considered to be roughly isotropic over the entire Fermi surface and the resistivity acquires the quadratic temperature dependence of a standard Fermi liquid. We have given a qualitative argument that explains this property as a consequence of umklapp scattering. Due to the strong requirements on the scattering geometry for umklapp processes, there is only a narrow region in band filling where the resistivity scaling is quadratic over a large range in temperature.

In a system with a specially nested Fermi surface, the numerically calculated resistivity exhibits a perfectly linear temperature dependence over more than an order of magnitude in temperature, down to the lowest temperature of our study. The explanation of this unconventional temperature scaling is based on a Brillouin-zone restriction of the effective scattering volume. It is tempting to believe that this scenario is of general importance for two-dimensional systems with experimentally observed linear resistance. A typical experimental example of systems that show a robust linear resistivity are optimally doped high-temperature cuprate superconductors. We have recently studied the temperature dependence of the resistivity of La$_{2-x}$Sr$_{x}$CuO$_{4}$ within our method\cite{buhmann_numerical_2013}. We believe that the form of the Fermi surface is a crucial aspect that should be considered in order to understand an experimentally observed linear resistivity of a correlated metal.

We have shown that two-particle scattering can not only lead to unconventional scaling, but its interplay with impurity scattering can also unveil intriguing properties. A simple example is a minimum in the resistivity as a function of the temperature. This unconventional behavior can be found for a strongly energy-dependent density of states as a consequence of anisotropic quasiparticle velocities and strong impurity scattering. Our simulation has illustrated the possibility of a resistivity minimum in the vicinity of a van Hove singularity. It should be stressed that the divergence of the van Hove singularity is not a crucial requirement for the effect; a strongly energy-dependent density of states is sufficient. This specific scenario illustrates how remarkable and unconventional transport properties can arise in generic metals and that such properties can only be understood by taking the angular and radial degrees of freedom of the collision integral into account.

We would like to note that Deng \emph{et al}. \cite{deng_how_2013} studied another mechanism to alter the generic quadratic temperature dependence of the resistivity. They considered the loss of well defined quasiparticle excitations within dynamical mean field theory and they identified two separated temperature scales. In their study, they showed that the scaling of the transport quantities in general is qualitatively changed across these temperature scales. In comparison to their study, we only consider the low-temperature regime to which Deng \emph{et al}. refer to as the Fermi-liquid regime, below both of their temperature scales. In our paper, we have argued that one can also find deviations from the predictions of Fermi-liquid theory in this regime.
 
Our study highlights the effect that details of the band structure can have on simple transport properties. A textbook-like quadratic temperature dependence of the resistivity from quasiparticle interactions, is often not observed experimentally. This study gives an understanding of how deviations from the standard Fermi-liquid scaling appear naturally even in simple systems. Such deviations are typically found at temperatures above the ultra-low-temperature limit, but under special conditions the unconventional behavior can extend down to zero temperature. 

We conclude that unconventional transport properties are not sufficient to characterize a system as a non-Fermi-liquid, in general. We have demonstrated through many examples that deviations from the standard transport behavior can arise within our model that classifies as a simple Fermi liquid with quasiparticles. However, given a good understanding of the quasiparticle states, our method yields precise predictions of the transport properties and, thus, might give insights to experimental observations that deviate from the predictions of standard Fermi-liquid theory.

\begin{acknowledgements}

The author is grateful to M. Ossadnik, T.M. Rice and M. Sigrist for many helpful discussions. This study was supported by the Swiss Nationalfonds, the NCCR MaNEP, the HITTEC project of the Competence Center Energy \& Mobility and the Sinergia TEO. 

\end{acknowledgements}

\end{document}